%% file: manuscript.tex
\documentclass{tex-library/jfm}

\shorttitle{Side-heated Rayleigh-B\'enard Convection}
\shortauthor{J. M. Huang and J. Zhang}

\title{Side-heated Rayleigh-B\'enard Convection}

\author{Jinzi Mac Huang\aff{1,2}\corresp{\email{machuang@nyu.edu}} and Jun Zhang\aff{1,2,3}\corresp{\email{jz11@nyu.edu}}}

\affiliation{
\aff{1}NYU-ECNU Institute of Physics and Institute of Mathematical Sciences, New York University Shanghai, Shanghai, China
\aff{2}Applied Math Lab, Courant Institute, New York University, New York, USA
\aff{3}Department of Physics, New York University, New York,
}

\include{preamble}

\begin{document}

\maketitle

\begin{abstract}
Unlike solids, the heat transfer in fluids can be greatly enhanced due to the presence of convection. Under gravity, an unevenly distributed temperature field results in differences in buoyancy, driving fluid motion that is seen in \RBC{} (RBC). In RBC, the overall heat flux is found to have a power-law dependence on the imposed temperature difference, with enhanced heat transfer much beyond thermal conduction. In a bounded domain of fluid such as a cube, how RBC responds to thermal perturbations from the vertical sidewall is not clear. Will sidewall heating or cooling modify flow circulation and heat transfer? We address these questions experimentally by adding heat to one side of the RBC. Through careful flow, temperature, and heat flux measurements, the effects of adding side heating to RBC are examined and analyzed, where a further enhancement of flow circulation and heat transfer is observed. Our results also point to a direct and simple control of the classical RBC system, allowing further manipulation and control of thermal convection through sidewall conditions.
\end{abstract}

\begin{keywords}
Authors should not enter keywords on the manuscript, as these must be chosen by the author during the online submission process and will then be added during the typesetting process (see http://journals.cambridge.org/data/\linebreak[3]relatedlink/jfm-\linebreak[3]keywords.pdf for the full list)
\end{keywords}

\section{Introduction}
\RBC{} (RBC) has been extensively studied in the past few decades \citep{tilgner1993temperature, belmonte1994temperature, niemela2000turbulent, funfschilling2005heat, ahlers2009heat, doi:10.1146/annurev.fluid.010908.165152}, for its vastly broad applications in geophysics \citep{zhong2005thermal, meakin2010geological, huang_zhong_zhang_mertz_2018,Wang2023,Huang2024}, solar physics \citep{nordlund1985solar,stein1989topology,DudokdeWit2020}, atmospheric science \citep{emanuel1994atmospheric,salmon1998lectures}, and ocean dynamics \citep{jorgensen1989vertical,vallis2006atmospheric}. In RBC, a domain of fluid is heated from below and cooled from above, so the near-boundary fluid is either warmed up or cooled off, leading to thermal convection. As the the warm and cold fluids move between the top and bottom plates, they effectively create an upward heat flux. At a fixed bottom-to-top temperature difference $\dT$, this convective heat transfer rate $Q_{conv}$ is much greater than that which would pass through a quiescent fluid body, $Q_{cond}$, which is the so-called conductive heat transfer rate. Defined as the ratio between the convective and conductive heat transfer rates, the Nusselt number $\Nu{} = Q_{conv}/ Q_{cond}$ measures the heat-passing capability in RBC, which depends on imposed control parameters such as the Rayleigh number, $\Ra{} =  \alpha g L^3  \dT  / (\kappa \nu)$, and the Prandtl number, $\Pr{} =  \nu/\kappa$. Here,  $g, L, \alpha,  \kappa, \nu$ are the acceleration due to gravity, fluid depth, thermal expansion coefficient, thermal diffusivity, and kinematic viscosity, respectively.

Once a fluid and its containing geometry are given, the heat flux is uniquely determined by the imposed temperature difference $\dT$. In particular, the famous scaling relationship $\Nu{}\propto \Ra^{\beta}$ has been observed in the range from $\Ra{}=10^6$ to $\Ra{} = 10^{14}$ \citep{niemela2000turbulent, funfschilling2005heat}, despite local deviations. Grossmann and Lohse \citep{grossmann2000scaling, ahlers2009heat, grossmann_lohse_2013} developed a theory -- the GL theory -- that incorporates the heat transfer contributed by both the thermal boundary layers and the bulk, revealing a $\Nu{}-\Ra{}$ relationship that has been found to be consistent with experiments in a large range of parameters.

At a given Rayleigh number, a controllable $\Nu{}$ is often desired; anomalies that deviate from the usual scaling laws found in nature also demand explanations \citep{barry1992mountain}. In the past, there have been many attempts to modify the $\Nu{}-\Ra{}$ dependency. For example, surface roughness was added to the boundary which modifies the boundary layer structure and enhances the Nusselt number \citep{PhysRevLett.81.987, PhysRevLett.120.044501}. Adding rotation to the entire fluid changes the bulk flow structure \citep{PhysRevLett.102.044502, PhysRevLett.103.024503, zhong_ahlers_2010} and increases the Nusselt number at moderate rotation rate. Stronger confinement on the convective fluid increases the flow structure coherence \citep{xia1997turbulent, chong_xia_2016, huang_xia_2016}, and leads to a higher Nusselt number. Bao et al. reported a strong $\Nu{}$ enhancement when vertical partition walls are inserted in the bulk fluid \citep{FLM:10030980}. There, the convective fluid self-organizes and circulates around these partitions, intruding into the thermal boundary layers and leading to increased heat flux.

Inspired by the fact that large-scale circulation can be induced by an imposed horizontal temperature gradient \citep{PhysRevE.51.5681}, we have demonstrated that the addition of horizontal flux can greatly enhance the large-scale circulation and heat transfer in RBC \citep{Huang2022}. In that work, a pair of heating-cooling vertical walls is added so a net horizontal flux can flow through the RBC cell. While such a study is numerically feasible, it is difficult to add and remove the same amount of heat experimentally. We thus focus only on the side-heating effect here, and experimentally demonstrate how an added horizontal flux can change the dynamical and thermal properties of RBC.

\begin{figure}
\centerline{\includegraphics[width=3.6in]{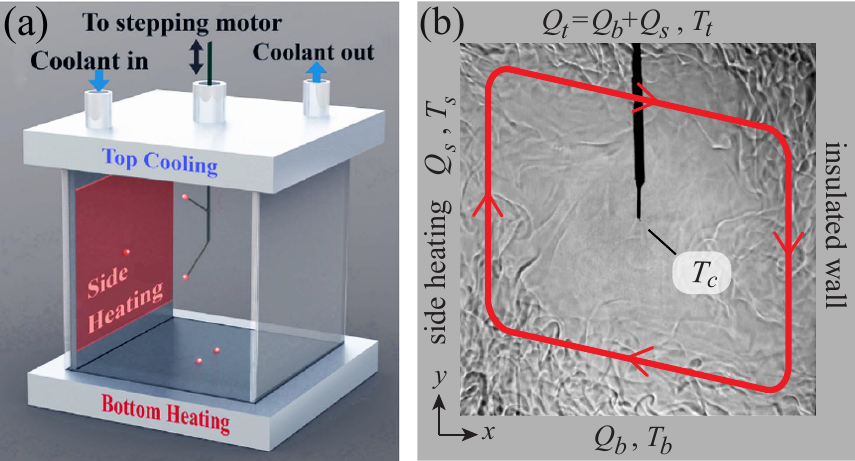}}
  \caption{Cubical thermal convection cell with sidewall heating is filled with water as the working fluid. (a) Schematic showing the convection cell and the side heater. Thermistors (red dots) are distributed to measure local temperature at various locations. (b) Shadowgraph visualizing the density difference and thus the flow structures, with bottom heating power $Q_b = 100$ W and side-heating power $Q_s = 40$ W. Heat injected from the bottom and the side exits the fluid through the top cooling plate, while the clockwise large-scale circulation is enhanced and dictated by the side heating. }
\label{fig1}
\end{figure}

In this study, a classical RBC system in a cubical domain is perturbed by a heat flux injected from one of its four vertical sides as shown in \cref{fig1}(a). During each experiment, the top cooling temperature $T_t$ is fixed, while $\Ra{}$ and $\Nu{}$ are measured as functions of the side and bottom heating powers $Q_s$ and $Q_b$. To the best of our knowledge, this unconventional configuration has not been studied in the past, despite many experiments having explored configurations such as sidewall heating-cooling without bottom heating \citep{PhysRevE.51.5681}, horizontal convection \citep{miller1968thermally, wang2005experimental,hughes2008horizontal}, and different sidewall boundary conditions \citep{FLM:9162434}.

In what follows, we will introduce the experimental setup in \cref{exp}, and show the main results regarding the flow patterns and heat transfer properties of the side-heated RBC in \cref{results}. And finally in \cref{discussion}, we will summarize our results and outline future research plans.

\section{Experimental setup}
\label{exp}

As shown in \cref{fig1}(a), our experiments are carried out in a cubical cell of $L = 20$ cm on each side. The top and bottom plates are made of anodized aluminum that ensures high thermal conductivity and uniform temperature distribution within each plate. In the top plate, coolant water is circulated at a constant temperature $40  ^{\circ}$C that is regulated by a water circulator. An electrical film heater is embedded in the bottom plate, providing bottom heating to the fluid. To ensure temperature uniformity, both the top and bottom plates are 2.54 cm thick and made of aluminum. Another film heater covered by an aluminum sheet, 2 mm in thickness and in contact with the fluid, is attached to one side of the vertical walls. The covering aluminum sheet is intended to improve temperature uniformity, whose top and bottom edges are cut away by 5 mm in order to prevent a ``short circuit'' of heat flowing from/to the boundary layers. During all experiments excluding the flow visualization, insulating materials cover the RBC cell to minimize heat leak. Local temperature is measured by thermistors distributed in the convection cell. As shown in \cref{fig1}(a), two thermistors measuring the bottom temperature are embedded 0.5 mm below the bottom surface. One thermistor is located at the center of the side heater to monitor the side temperature, and two thermistors are mounted on a support that can traverse vertically to measure the temperature profile of the fluid.

We perform all temperature and flow measurements at each $(Q_b,Q_s)$, while varying them independently in the range of 0 to 120 W. While the heat mainly transfers through the fluid, there are three possible heat leaks that may affect our measurements: 1. the conduction beneath the bottom plate; 2. the vertical conduction of heat through the side walls; 3. the horizontal heat leak from the side walls to the air. 

To minimize the conductive heat leak, the bottom plate is placed on top of layers of insulating foam, beneath which a compensating heater is installed so the temperature gradient in the foam can be monitored and minimized by a PID algorithm. The sidewalls are made of $5$ mm thick acrylic, whose thermal conductivity is around $0.2$ W/(m$\cdot$K). Considering the typical temperature difference between top and bottom is about $10^\circ$C, the vertical conducting power through the side walls is estimated as $0.04$ W, much smaller than the convective power carried by the fluid.

The main contribution to heat leakage is in fact the heat transfer between the side walls and the surrounding air. Typically, the convection cell operates at a bulk temperature of $40^\circ$C, and the surrounding air is at the room temperature of $25^\circ$C. The convection cell, when operating, is covered by many layers of aluminum-coated bubble wrap whose total thickness is on the order of $10$ centimeters and conductivity is around $0.02$ W/(m$\cdot$K). This allows us to estimate the total side leaking power as $0.5$ W, which is much smaller than the minimum bottom heating power ($10$ W) used in the study. Thus, the maximum error we may have in measuring Nu is 5\%, and this error decays inversely with $Q_b$.

Whenever a control parameter is changed, the system runs for 4 hours to reach a dynamical equilibrium, and each measurement takes another 4 hours to collect data. With degassed water as the working fluid, the system works at Rayleigh number in the range of $3\times10^8$ to $5\times10^9$, and Prandtl number in the range of $3.4$ to $4.1$. At this $\Ra{}$, the convection is turbulent and the large-scale circulation develops spontaneously.

Shown in \cref{fig1}(b), a shadowgraph of the convecting fluid reveals local density differences and thus provides a visualization of the convective flows. A convex lens with focal length 20 cm converts light from a point source to near-parallel light, which passes through the convection cell and casts a shadow on a translucent screen. Bulk flow speed $U$ can be inferred by tracing the thermal plumes at a given region, which we choose to be 1 cm from the center of the heated sidewall. This flow speed $U$ represents the strength of the large-scale circulation. Typically, trajectories of 40 thermal plumes are timed and metered during each experiment, so both the mean flow speed and its standard deviation can be calculated.

\section{Results}
\label{results}
\subsection{Enhanced large-scale circulation and modified $\Nu{}-\Ra{}$ relationship}
\label{results-part1}

\begin{figure}
  \centerline{\includegraphics[width=4.2in]{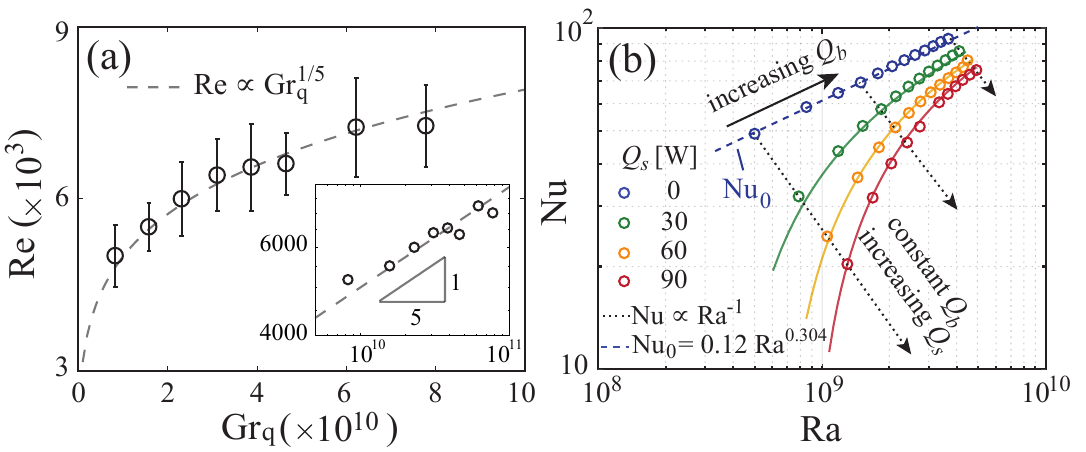}}
  \caption{ Flow speed and heat transfer measured by Reynolds and Nusselt numbers. (a) Plume speed increases with side-heating power. With bottom heating fixed at 100 W, Reynolds number $\Reyy{} =UL/\nu$ measured at 1 cm from the sidewall center shows a monotonic dependence on the power Grashof number $\Gr_q$, consistent with the power-law scaling $\Reyy{}\propto \Gr{}_q^{1/5}$ from the boundary layer theory \citep{schlichting1960boundary}. \textit{Inset}: Logarithmic-scale plot of Reynolds number versus power Grashof number. (b) Nusselt number $\Nu{}$ measured at various values of $\Ra{}$ and side-heating power $Q_s$. The bottom heating power is $Q_b = 10,20,\dots120$ W. Without side-heating, $\Nu{}$ shows a classical power-law dependence on $\Ra{}$. Fixing the bottom power, $\Nu{}$ is observed to decrease with increasing side power, while $\Nu{}\propto \Ra{}^{-1}$. Colored solid curves are the $\Nu{}-\Ra{}$ curve at each fixed $Q_s$ predicted by a theory introduced in \cref{results-temp}.}
\label{fig2}
\end{figure}

We expect the addition of side heating to enhance the large-scale circulation, and perhaps also to increase the vertical heat transfer rate. Shown in \cref{fig1}(b), the fluid near the sidewall is heated and becomes lighter, thus forming an upwelling jet that feeds a clockwise large-scale circulation. Therefore, the thermal energy provided by sidewall heating is partly  converted into the fluid kinetic energy that enhances the large-scale circulation. 

With a constant bottom heating power $Q_b = 100$ W and side-heating power in the range of $Q_s = 10-100$ W, the flow speed can be nondimensionalized as the Reynolds number $\Reyy{} = UL/\nu$ shown in \cref{fig2}(a). In \cref{fig2}(a), the sidewall power is also nondimensionalized as the \textit{power} Grashof number $\Gr{}_q = \alpha gL^2 Q_s/(\lambda \nu^2)$, which is a measure of the relative strength between buoyancy and viscous effects \citep{schlichting1960boundary}. Here, $\lambda$ is the fluid thermal conductivity. In comparison with the Grashof number $\Gr{} = \alpha g L^3 (T_s - T_\infty)/\nu^2$, the power Grashof number is defined directly through the input heating power $Q_s$ instead of the sidewall temperature difference $T_s - T_\infty$. Far field temperature $T_\infty$ in our case can be taken as the bulk temperature.

The monotonic increase of $\Reyy{}$ with $\Gr{}_q$ shows that the bulk flow speed is indeed accelerated by the side-heating power. From the boundary layer theory \citep{schlichting1960boundary}, the buoyancy-driven flow speed near a vertical wall scales with the Grashof number $\Gr{}$ as $U \propto \Gr{}^{1/4}$. Using the relationship between the two Grashof numbers \citep{schlichting1960boundary}, namely $\Gr{} \propto \Gr{}_q^{4/5}$, we have the scaling law $\Reyy{}\propto \Gr{}^{1/4} \propto \Gr{}_q^{1/5}$. This is found to be consistent with our experimental data, as shown in \cref{fig2}(a). As $Q_s$ reduces to 0, flow velocity drops to the conventional \RBC{} value, which corresponds to $\Reyy{}\sim 1000$ in our range of $\Ra{}$ \citep{ahlers2009heat}. Without side heating, the direction of large-scale circulation becomes arbitrary but stays along one of the vertical diagonal planes of the convection cell.

The Nusselt number in the classical cubical RBC is $\Nu{} = Q_b / (\lambda \dT L)$, and \cref{fig2}(b) shows the Nusselt number measured at various values of $Q_s$ and $Q_b$. With sidewall heating added, the bulk circulation is enhanced as expected, but the bottom temperature $T_b$ and the resulting $\dT = T_b - T_t$ are also found to increase. By definition, $\Ra{} \propto \dT$ and $\Nu{} \propto  1 / \dT$, so an increased $\dT$ due to side-heating causes $\Ra{}$ to increase and $\Nu{}$ to decrease. Moreover, with $Q_b$ fixed at a constant, the product $\Nu{}\cdot\Ra{}$ is independent of $\dT$. Thus, the dotted lines in \cref{fig2}(b) have a common slope -1 in the log-log scale.

As a reference, the $\Nu{}-\Ra{}$ measurement without sidewall heating, $\Nu{}|_{Q_s = 0} = \Nu{}_0$ in \cref{fig2}(b), agrees well with previous experiments \citep{funfschilling2005heat, ahlers2009heat} and follows a power-law scaling $\Nu{} = 0.12 \Ra{}^{0.304}$ that is consistent with the GL theory. When $Q_s \neq 0$, the observed data deviate from the power-law relation -- a simple scaling relationship is absent, particularly at large $Q_s$.

\subsection{Temperature measurement explaining the $\Nu{}-\Ra{}$ scaling}
\label{results-temp}
\begin{figure*}
\centerline{\includegraphics[width=5in]{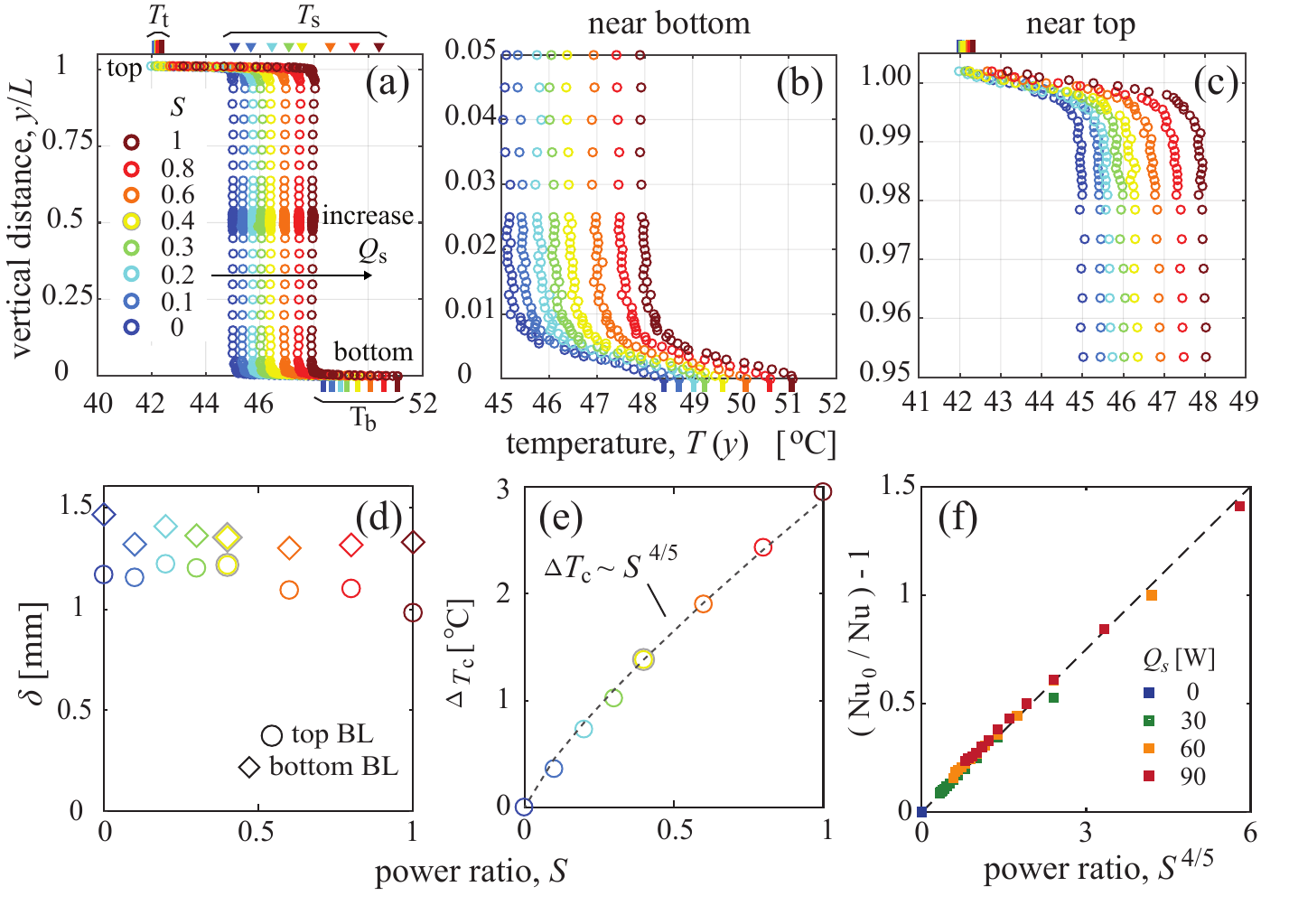}}
  \caption{Vertical temperature profiles measured at $Q_b = 100$ W and various values of $Q_s$, and along the center vertical line as shown in \cref{fig1}(a). (a) Temperature at different height. The bulk and bottom temperatures are found to increase with the heating power ratio $S = Q_s/Q_b$. Top, bottom and side temperatures ($T_t, T_b$ and $T_s$) are marked with ticks and triangles. (b) Zoom-in profiles of the bottom thermal boundary layer. $T_b$ increases with the bulk temperature, while the temperature gradient stays nearly constant. (c) Zoom-in profiles of the top thermal boundary layer. With top temperature fixed, increased bulk temperature leads to a higher temperature gradient within the boundary layer. (d) Thermal boundary layer thicknesses $\delta$ of the top and bottom boundary layers. (e) With $Q_b$ fixed, the increase of bulk temperature $\dT_c = T_c|_S - T_c|_{S=0}$ has a $4/5$ power-law scaling relationship to the side-heating power $Q_s$. (f) Nusselt number as a function of power ratio $S$, $\Nu{}_0/\Nu{} = 1+ 0.25 S^{4/5}$. [(a)-(e) share the same color scheme; (f) uses the color scheme and definition of $\Nu{}$ and $\Nu_0$ as in \cref{fig2}(b).]}
\label{fig3}
\end{figure*}

To investigate the modified $\Nu-\Ra{}$ relationship due to sidewall heating, the vertical temperature distribution is examined along the central vertical axis of the RBC cell shown in \cref{fig1}(a).  \Cref{fig3}(a) shows several temperature profiles measured at the same bottom power $Q_b = 100$ W but different side power $Q_s$, where we have defined the heating power ratio as $S = Q_s/Q_b$.  On each curve, temperature changes linearly within thermal boundary layers, through which heat transfers by conduction. In the bulk, strong turbulent mixing makes the bulk temperature $T_c$ nearly uniform \citep{belmonte1994temperature, ahlers2009heat}, which, together with the bottom temperature $T_b$, increases with $S$.

Zooming in near the bottom,  \cref{fig3}(b) shows that the bottom temperature $T_b$ increases with $Q_s$ while the temperature gradient $\partial T / \partial y$ stays roughly constant within the boundary layer. This is due to the fact that  $\partial T / \partial y = -Q_b / (\lambda L^2)$ and the bottom power $Q_b$ is fixed at a constant. The bottom thermal boundary layer thickness $\delta_b = 1.4 \pm 0.1$ mm or $\delta_b / L = (0.70 \pm 0.05 )\%$ is roughly constant for varying $S$ as shown in  \cref{fig3}(d). This, together with a constant $\partial T/\partial y \sim \dT_b / \delta_b$, indicates the temperature difference between the bottom and the bulk $\dT_b = T_b - T_c $ is not changing with $S$ either, as shown in \cref{fig3}(b). \Cref{fig3}(c) shows the zoom-in temperature profiles near the top plate. The addition of side heating $Q_s$, which has to leave the system through the top plate, leads to a higher temperature gradient $\partial T / \partial y = -(Q_b+Q_s)/(\lambda L^2)$ in the top thermal boundary layer.  In our experiment, the top thermal boundary layer thickness is $\delta_t = 1.1 \pm 0.1$ mm or $\delta_t / L = (0.55 \pm 0.05)\%$, as shown in \cref{fig3}(d).

Interestingly, the top thermal boundary layer is thinner than the bottom in \cref{fig3}(d), even when $S=0$. This is the opposite of the classic RBC, where the non-Boussinesq effect of working fluid makes the bulk warmer and the bottom thermal boundary layer thinner \citep{Zhang1997}. We have confirmed that this anomaly is a consequence of adding the aluminum side heater, which both conducts heat and modifies the flow structure even when the heating is off. For $S \neq 0$, the boundary layer thickness also varies in space, so the temperature distribution shown in \cref{fig3} only reflects a local profile covered by the range of moving thermistors. To better understand the temperature and flow distributions in the side-heated RBC, we are currently working on a numerical study that may provide further insight.

As the only variable in $\Nu = Q_b / (\lambda \dT L)$ is $\dT$ when holding $Q_b$ constant, understanding how $\dT$ depends on $S$ can directly explain how $\Nu{}$ depends on $S$. From the data shown in \cref{fig3}(b), the bottom to bulk temperature difference stays unchanged while the bulk temperature increases with $S$. Therefore, the increase of bulk temperature $\dT_c (S)= T_c|_S - T_c|_{S=0}$ directly contributes to an increased bottom-top temperature difference $\dT$. Shown in \cref{fig3}(e), the bulk temperature change has a scaling $\dT_c \propto S^{4/5}$, which is a direct consequence of the scaling $\Gr{} \propto \Gr{}_q^{4/5}$ discussed earlier in \cref{results-part1}.

We therefore estimate $\dT = (1 + \gamma S^{4/5}) \dT |_{S=0}$, where $\gamma$ is a positive constant. Substituting this into the definition of Nusselt number, we have $\Nu{}_0/\Nu{}(S) = \dT/\dT|_{S=0} = 1 + \gamma S^{4/5}$. Plotting $\Nu{}_0/\Nu{} - 1$ against $S^{4/5}$, all the data points in \cref{fig2}(b) land on a straight line as shown in \cref{fig3}(f), whose slope is $\gamma \approx 0.25$.  We can also apply this relation to $S = Q_s/Q_b$ with fixed $Q_s$ and varying $Q_b$, which is shown as the $\Ra{}-\Nu{}$ curve for each $Q_s$ in \cref{fig2}(b).

The above analysis provides a relationship between $\Nu{}$ and $S$, thus suggesting the heat transfer in RBC can be controlled by simply changing $S$. However an obvious question still remains: why does a stronger large-scale circulation shown in \cref{fig2}(a) lead to a reduced $\Nu{}$ for side-heated RBC? In the next section, we will measure the true heat flux by redefining the Nusselt number, and show how large-scale circulation enhances the heat transfer of side-heated RBC. 

\subsection{Redefined Nusselt number and its enhancement}
\label{redef}

In this section, we redefine the Nusselt number by counting {\it all} heat inputs to the RBC system as the total convective heat $Q^*_{conv}$, and compare it with the {\it overall} conductive heat $Q^*_{cond}$ associated with the wall temperature distribution. The resulting analysis will show that the newly defined Nusselt number $\Nu{}^* = Q^*_{conv}  / Q^*_{cond}$ indeed increases with $S$.

\begin{figure}
  \centerline{\includegraphics[width=4.5in]{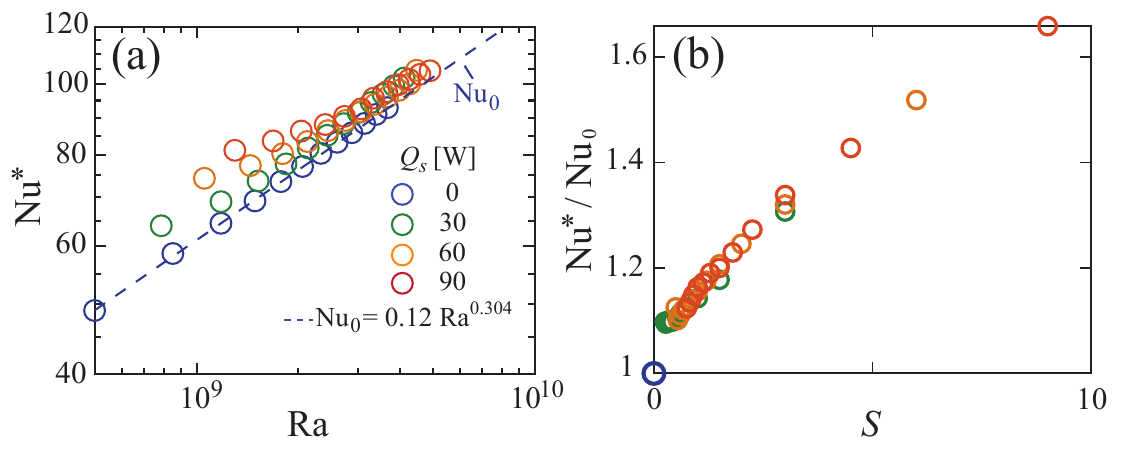}}
  \caption{ Redefined Nusselt number $\Nu{}^*$ as the ratio of total convective heat to total conductive heat. (a) With redefinition, data shown on \cref{fig2}(b) are now above the $\Ra{}-\Nu_0$ curve when $Q_s>0$, showing an enhancement of heat transfer. (b) The ratio $\Nu{}^*/$Nu$_0$ increases monotonically with the heating power ratio $S = Q_s/Q_b$,  reaching the highest enhancement of $66\%$ at $S = 9$ in the experiment. [(a)-(b) share the same color scheme as in \cref{fig2}(b)]}
\label{fig4}
\end{figure}

As we have argued above, the sum of the side and bottom heat has to flow out through the top plate. In terms of power, the total convective power through the cell is $Q^*_{conv} = Q_t = Q_b + Q_s$, where $Q_t,\ Q_b$, and $Q_s$ represent the magnitude of power passing through the top, bottom, and side, respectively. The side heating also leads to a side temperature $T_s > T_c> T_t$ and it contributes to the total conductive heat flowing through the top plate. Hence $Q_{cond}^*$ has to be determined by solving a steady state heat equation with $T_b,\ T_t$, and $T_s$ satisfying Dirichlet boundary condition for the bottom, top, and one of the sidewalls, and $\partial_n T = 0$ for the other 3 sidewalls. To prevent short-circuiting, the heating sidewall is separated into three regions, one central region with constant temperature $T_s$ and two stripes on its top and bottom with adiabatic boundary conditions, whose size matches the experimental configuration of $5:190:5 \mbox{ mm}$. The steady-state heat equation $\partial_x^2 T + \partial_y^2 T = 0$, together with these mixed boundary conditions, can be solved numerically. Here we use a finite elements package COMSOL Multiphysics with MATLAB Livelink and compute the conductive heat in 2D, as we assume no significant structural change in the third direction. The overall conductive heat transfer rate, computed along the top boundary, is thus determined as $Q^*_{cond} = - \lambda L \int_0^{L} \frac{\partial T}{\partial y} (x, L) dx$. 

As shown in \cref{fig4}(a), the redefined $\Nu{}^*=Q^*_{conv}/Q^*_{cond}$ for each $\Ra{}$ and $S>0$ lies above the unperturbed $\Ra{}-\Nu{}_0$ curve [dashed line in \cref{fig4}(a)], indicating side heating indeed results in a higher heat transfer rate. At large $\Ra{}$ (small $S$), contribution from the bottom heating $Q_b$ dominates the overall heat transfer, so the classical $\Nu{}_0-\Ra{}$ relationship becomes the asymptotic limit. At small $\Ra{}$ (large $S$), however, the redefined $\Nu{}^*$ deviates from the conventional power-law scaling. The degree of this enhancement, measured as $\Nu{}^*/\Nu{}_0$ in \cref{fig4}(b), increases monotonically with $S$ and reaches an enhancement of $66 \%$ at $S=9$ -- the highest $S$ reached in our experiments. We attribute this $\Nu{}^*$ boost to the strengthening of large-scale circulation due to side heating, which enhances the bulk mixing and perhaps thins the boundary layers [\cref{fig3}(d)].

\section{Discussions}
\label{discussion}

Through laboratory experiments, we have investigated the effects of adding side heating to the classical RBC system. In particular, we have found an empirical law $\Nu{}(S) = \Nu{}_0 (1 + \gamma S^{4/5})^{-1}$, which deviates from the classical $\Nu{}-\Ra{}$ scaling. We further show that a redefinition of $\Nu{}$ is required to capture the unconventional boundary conditions in our study, and the redefined $\Nu{}^*$ increases monotonically with the side-heating power, thus allowing for a direct control to the RBC.

Such controllability was also demonstrated in \cite{Huang2022}, where adjusting the horizontal heat flux leads to a response of $\Nu{}$ much similar to \cref{fig4}(a). In that work, $\Nu{}$ also increases monotonically with $S$, with the highest relative enhancement achieved near the critical Rayleigh number. In the limit of $S\to 0$, the horizontal flux becomes negligible compared to the vertical one, so $\Nu{}\to\Nu{}_0$ and the RBC converges to the classic configuration asymptotically. 
On the other hand, taking $S\gg 1$ would bring the system closer to the configuration of \cite{PhysRevE.51.5681}, which is a RBC that is turned $90^\circ$ to the side.

Thus, both the present experiment and the previous numerical simulation \citep{Huang2022} provide a simple means of controlling the RBC -- adjust the horizontal heat flux and the vertical flux responds accordingly. In electronics, current is a direct analogy to the heat flux in thermal dynamics. Therefore our side-heated RBC functions much like an electronic NPN transistor, where the current flowing between the collector and the emitter can be controlled by adding a current to the base \citep{note1}. 

But unlike the electronic transistors, the response of our ``thermal transistor" is slow, usually taking hours to reach dynamical equilibrium after adjusting the side-heating power. In \cite{Huang2022}, we have experimented with a time-dependent perturbation and identified a relaxation time for the system to reach dynamical equilibrium. How does the side-heated RBC respond to a change of side-heating power? Our observation is that bulk quantities like $\Reyy{}$ and $\Nu{}$ relax to their equilibrium values exponentially, with a relaxation timescale of hours. 

With water as the working fluid, the $\Ra{}$ range for the present study is limited. We plan to conduct numerical studies and also work with other fluids to extend this range of $\Ra{}$. At very high $\Ra{}$, the RBC will transition into its ultimate regime \citep{10.1063/PT.3.5341}, and a noticeable perturbation from the side heating would require a significantly higher power in that case. However at very low $\Ra{}$, the fluid may stay in place if $\Ra{}$ is below its critical value-- side heating could thereby bring motion to an otherwise motionless fluid. A more detailed perturbation study is currently underway, and we hope to better understand the dynamical interplay between the boundary conditions and the bulk fluid through such fluid-structure-interaction investigations.



\bibliographystyle{tex-library/jfm}
\bibliography{manuscript}

\end{document}

%% file: preamble.tex
\usepackage{graphicx, amsmath, amssymb, amsfonts, mathtools, mathrsfs, color}
\usepackage{comment, todonotes}
\usepackage{soul}
\usepackage{dcolumn}

\usepackage{xcolor}
\usepackage[capitalise]{cleveref}
\usepackage[normalem]{ulem}
\usepackage{bm}
\usepackage{url}

\newcommand{\bvec}[1]{\ensuremath{\boldsymbol{#1}}}

\newcommand{\RBC}{Rayleigh-B\'enard convection}

\newcommand{\Ra}{\text{Ra}}
\newcommand{\Reyy}{\text{Re}}
\newcommand{\Nu}{\text{Nu}}
\newcommand{\Gr}{\text{Gr}}
\newcommand{\dT}{\Delta T}

\renewcommand{\eth}{\bvec{e_\theta}}

